\documentclass[doublecol]{epl2} 

\usepackage{graphicx}% Include figure files

\usepackage{bm}% bold math
\usepackage{amssymb}
\usepackage{amsmath}
\usepackage{amstext}
\usepackage{latexsym}
\usepackage[colorlinks,linkcolor=red]{hyperref}
\def\x{{\rm x}}

\def\la{\langle}
\def\ra{\rangle}
\def\om{\omega}

\newcommand{\beq}{\begin{equation}}
\newcommand{\eeq}{\end{equation}}
\newcommand{\beqa}{\begin{eqnarray}}
\newcommand{\eeqa}{\end{eqnarray}}

\title{Fast frictionless dynamics as a toolbox for low-dimensional Bose-Einstein condensates}
\shorttitle{Fast frictionless dynamics in low-dimensional Bose-Einstein condensates}

\author{A. del Campo\inst{1,2}}

\institute{                    
  \inst{1} Institut f{\"u}r Theoretische Physik, Leibniz Universit\"at Hannover - Appelstrasse 2, D-30167 Hannover, Germany\\
  \inst{2} Institut f{\"u}r Theoretische Physik, Universit{\"a}t Ulm - Albert-Einstein Allee 11, D-89069 Ulm, Germany
}


\abstract{
A method is proposed to implement a fast frictionless dynamics in a low-dimensional Bose-Einstein condensate by engineering the time-dependence of the transverse confining potential in a highly anisotropic trap.
The method exploits the inversion of the dynamical self-similar scaling law in the radial degrees of freedom.
We discuss the application of the method to preserve short-range correlations in time of flight experiments, the implementation of nearly-sudden quenches of non-linear interactions, and its power to assist self-similar dynamics in quasi-one dimensional condensates.
}

\begin{document}
\maketitle

Counterintuitive as they are, implementations of fast frictionless dynamics (FFD) providing a shortcut to adiabaticity in quantum systems have recently been introduced theoretically \cite{chen10,Muga,Stefanatos} and  demonstrated in the laboratory \cite{Labeyrie10,Labeyrie10b}.

In this Letter, FFD is exploited as a tool-box to manipulate and control low dimensional quantum gases. As a primary goal, we propose a method to tune the non-linearity of the effective low dimensional (1D and 2D) dynamics of an anisotropic Bose-Einstein condensate (BEC) by engineering the time-modulation of the transverse confinement. The possibility of attaining this goal by means of a multi-scale expansion method was recognised by Staliunas {\it et al.} for {\it slow} periodic modulations \cite{Staliunas04} and led to the observation of Faraday patterns in cigar-shaped BECs \cite{Engels07}.

Our method, not restricted by adiabaticity in the transverse dynamics, provides an alternative way to implement a variety of schemes for the generation, stabilisation and control of solitons, and the study of  related  non-linear matter-wave phenomena in BEC,  where it is often necessary to implement a time-dependent coupling constant \cite{solitons}.
Other situation where it can be applied is in the controlled expansion of BEC clouds \cite{KSS97} as in the simulation of cosmological analogues \cite{Barcelo}.
Moreover, we shall discuss how FFD can be exploited to implement nearly-sudden quenches of the mean-field non-linear interactions, preserve quantum correlations in time-of-flight and induce a self-similar expansion in interacting quasi-1D atomic cloud.

FFD allows to evolve from an initial state to a final one without exciting the system while doing it in a given time $\tau$ much smaller than that required for an adiabatic dynamics.  FFD relies on the inversion of dynamical scaling laws, which can often be exploited to described harmonically trapped ultracold gases, such as the Calogero-Sutherland model \cite{Sutherland98}, Tonks-Girardeau \cite{OS02,MG05}, strongly interacting mixtures \cite{MG07}, Lieb-Liniger gases \cite{BPG08}, Bose-Einstein condensates (BEC) \cite{CD96,KSS96,Muga}, including dipolar interactions \cite{dipolar}, and more general many-body quantum systems \cite{demler}.
In a harmonic trap with time-dependent frequency, the self-similar evolution of the single-particle states $\phi_n$ ($n=0,1,2,\dots$) follows from the well-known scaling law,
\beqa
\label{scaling}
\phi_n(x,t)=\phi_n(x/b(t),0)e^{imx^2\dot{b}/2b\hbar-iE_n\delta(t)/\hbar}/\sqrt{b(t)}
\eeqa
where the scaling factor $b=b(t)$ is the solution of the Ermakov differential equation
\beqa
\label{EPE}
\ddot{b}+\om^2(t)b=\om(0)^2/b^3,
\eeqa
satisfying the boundary conditions $b(0)=1$ and $\dot{b}(0)=0$, with $E_n=\hbar\om(0)(n+1/2)$, and $\delta(t)=\int_{0}^tdt'/b^2(t')$ \cite{invariants}. In particular, the probability density reads $|\phi_n(x,t)|^2=|\phi_n(x/b(t),0)|^2/b(t)$. The essence of FFD is to exploit the existence of the self-similar dynamics to force a desired trajectory $b(t)$ and invert Eq. (\ref{EPE}) to determine the required modulation of the control parameter $\om(t)$ \cite{chen10,Muga}.

\section{Effective one-dimensional time-dependent Gross-Pitaevskii equation} 

Our aim is to find a 1D effective non-linear Schr\"odinger equation  or Gross-Pitaevskii (GPE) equation for a BEC in a elongated trap in which the transverse confinement is modulated in time, ${\rm V^{ex}}(\x,t)={\rm V^{ex}}(z)+\frac{m}{2}[\om_x(t)^2x^2+\om_y(t)^2y^2]$.  We will do so exploiting the scaling law in Eq. (\ref{scaling}).
We start with the 3D time-dependent GPE which governs the dynamics of the order parameter $\Psi=\Psi(\x,t)$
\beqa
\label{3DGPE}
i\hbar\frac{\partial\Psi}{\partial t}=\big[-\frac{\hbar^2}{2m}\Delta_{\x}+{\rm V^{ex}}(\x,t) +g_{\rm 3D}|\Psi|^2\big]\Psi,
\eeqa
with the normalisation condition $\int d\x|\Psi(\x,t)|^2=1$, and where $g_{\rm 3D}=\frac{4\pi\hbar^2Na}{m}$, $N$ is the number of atoms in the condensate of mass $m$, and $a$ the $s$-wave scattering length which determines the healing length $\xi=1/\sqrt{8\pi n|a|}$. It is convenient to introduce the characteristic length in each direction, $a_j=\sqrt{\hbar/(m\om_j(0))}$ for $j=x,y,z$, in terms of which the density scales as $n\sim N/\prod_j a_j$.
For tight transverse confinement ($\om_x\sim\om_y\gg \om_z$), the kinetic energy in the transverse direction governs over 2-body collisions. Whenever the dimensionality parameter $\epsilon^2=a_xa_y/\xi^2\sim N|a|/a_z\ll 1$, the transverse excitations are frozen and a dimensional reduction of the 3D GPE is possible \cite{PGMH98,CC02,BJM03}. The 3D GPE becomes a linear Schr\"odinger equation
for the radial degrees of freedom.
We next use the ansatz  $\Psi(\x,t) = \Phi_0(x,y,t)\psi(z,t)$ and consider the ground state condensate wave function in the transverse direction at $t=0$ to be that of an unperturbed 2D harmonic oscillator, $\Phi_0(x,y,t=0)$.
%
%\beqa
%\Phi_0(x,y,t=0)=\frac{1}{\sqrt{\pi a_x a_y}}
%\exp\left\{-\frac{1}{2}\big[\left(\frac{x}{a_x}\right)^2+\left(\frac{y}{a_y}\right)^2\big]\right\}.
%\nonumber
%\eeqa
%
 It follows from Eq. (\ref{scaling}) that  $|\Phi_0(x,y,t)|^2=|\Phi_0(x/b_x(t),y/b_y(t),t=0)|^2/(b_x(t)b_y(t))$.
%Multiplying the 3D GPE, Eq. (\ref{3DGPE}), by the complex conjugate of the transverse wavefunction $\Phi_0(x,y,t)^*$ and integrating over $x$ and $y$,
After dimensional reduction,
the following effective 1D GPE with a time-dependent non-linearity is derived, %\cite{note},
\beqa
i\hbar\frac{\partial\psi(z,t)}{\partial t}=
\big[H_z+g_{\rm 1D}(t)|\psi(z,t)|^2\big]\psi(z,t),
\eeqa
where $H_z=-\frac{\hbar^2}{2m}\frac{\partial^2}{\partial z^2}
+{\rm V^{ex}}(z)-\alpha(t)$, and the term $\alpha(t)=\sum_{j=x,y}\la H_j(t)\ra_0=\sum_{j=x,y}\hbar(4\om_0)^{-1}(\dot{b}_j^2+\om_j(t)^2b_j^2+\om_0^2/b_j^2)$ can be removed with a unitary transformation.
In the following, we focus our attention of the effective coupling constant defined as
\beqa
\label{effg1d}
g_{\rm 1D}(t)
\!=\!g_{\rm 3D}\!\!\iint\!\! dx dy |\Phi_0(x,y,t)|^4=\frac{2\hbar^2 N a}{\prod_{j=x,y}a_jb_j(t)},
\eeqa
satisfying the conditions
\beqa
\label{g0cond}
g_{\rm 1D}(0)=\frac{2\hbar^2 N a}{m a_xa_y}, \quad \dot{g}_{\rm 1D}(0)=0, \quad \ddot{g}_{\rm 1D}(0)=0
\eeqa
%The first two conditions warrant that $[\hat{I}_j(0),\hat{H}_j(0)]=0$ ($j=x,y$).
The third equality is optional, it follows from imposing the first two and the Ermakov equation with the continuity restriction $\om_{r=x,y}(t=0^+)=\om_{r=x,y}(t=0^-)$. Hence, it warrants a smooth modulation of the transverse trapping frequency and avoids abrupt changes at $t=0$. Nonetheless, abrupt changes of the control parameter are often exploited in bang-bang control methods \cite{Salomon}.

As pointed out by Olshanii, the transverse confinement can severely modify the scattering properties, and in particular, lead to a confinement-induced resonance of the form $g_{1D}\rightarrow g_{1D}(1-\mathcal{C}a/a_{\perp})^{-1}$, with $\mathcal{C}=1.4603\dots$ \cite{Olshanii98,BDZ08}. Eq. (\ref{effg1d}) holds far away from confinement induced resonances, i.e. $|a|\ll\{a_x,a_y\}$. 
In addition, low-dimensional BECs generally exhibit phase fluctuations which can lead to deviations from the mean-field \cite{PSW01,SF07}. 
%In a length scale of the axial size of the cloud, 
%they are reprenented by the parameter $\delta_L^2\propto (T/T_c)(\frac{\om_{\rho}}{N^4\om_z})^{19}$. 
%Its suppression is controlled by the parameter  which sets a bound of the possible changes in the trap aspect ratio. 
Even at zero-temperature, quantum fluctuations suppress the off-diagonal long-range order, in the sense that the reduced one-body density matrix 
$\rho_1(x,x')\rightarrow\chi(x)^*\chi(x')$ with $\chi(x)=\sqrt{n(x)}\exp(-\Gamma_D)$ for $|x-x'|\simeq R$ being $R$ the size of the system and $n(x)$ the local density. 
The quantum suppresion parameter for $D=1$ is given by
$\Gamma_1=\frac{2\sqrt{2}}{\pi}\left(\frac{\om_{\perp}}{\om_z}\right)\left(\frac{a}{R}\right)\ln\left(\frac{R}{\sigma_z}\right)$
where $\om_{\perp}^2=\om_x^2(0)+\om_y^2(0)$ \cite{HoMa99}.
% could be referred either to the initial or final state.
The applications discussed in the rest of the manuscript, 
involve the monotonic expansion of the cloud in the transverse direction ($\dot{b}(t>0)>0$) which diminishes the role of phase fluctuations. It will suffice for us to focus on situations where phase fluctuations are negligible in the initial state so as to prevent the formation of density ripples for $t>0$ \cite{ripples}, i.e. $\Gamma_1\ll 1$. However, should one be interest in increasing the effective interactions to a maximum value 
$g_{\rm 1D}(\tau)$, the condition for phase fluctuations to be negligible becomes $\Gamma_1 g_{\rm 1D}(\tau)/g_{\rm 1D}(0)\ll 1$. 
%. For this to be the case, using $\om_{\rho}^2(0)=\om_x^2(0)+\om_y^2(0)$, the parameter $\delta_L^2\approx (T/T_c)(\frac{a^6m^3\om_{\rho}^{22}(0)}{N^4\om_z^{19}(0)})^{1/15}\ll 1$ \cite{PSW01}. 
We note as well that the scaling function $b(t)>0$, so that it is not possible to change the sign of the interactions as with Feschbach or confinement-induced resonances \cite{BDZ08}.  This technique, is therefore restricted to tune the amplitude of the coupling constant. 

Let us now consider an isotropic transverse confinement such that $\om_{\perp}(t)=\om_x(t)=\om_y(t)$ ($b=b_x=b_y$), and assume we are interested in a given time dependence $g_{\rm 1D}(t)=g_{\rm 1D}(0)/b(t)^2$. This can be engineered by a trap modulation of the transverse trapping frequency given by
\beqa
\label{1dom2}
\om_{\perp}^2(t)=\om_{\perp}(0)^2\left(\frac{g_{\rm 1D}}{g_{\rm 1D}(0)}\right)^2
+\frac{1}{2}\frac{\ddot{g}_{\rm 1D}}{g_{\rm 1D}}
-\frac{3}{4}\left(\frac{\dot{g}_{\rm 1D}}{g_{\rm 1D}}\right)^2,
\eeqa
where $g_{\rm 1D}=g_{\rm 1D}(t)$.
Note that this expression is not positive-definite, and $\om_{\perp}^2(t)$ might involve imaginary frequencies as discussed below.
It is straightforward to verify that for ballistic expansion along the transverse degrees of freedom, which leads to a polynomially decaying non-linearity $g_{\rm 1D}(t)=g_{\rm 1D}(0)/[1+\om_{\perp}^2(0)t^2]$ \cite{RDM05}, the required transverse frequency is indeed $\om_{\perp}^2(t>0)=0$.
Moreover, the existence of the invariant of motion \cite{chen10} associated with the Ermakov equation underlying the derivation of Eq. (\ref{1dom2}) allows to drive the transverse frequency without fulfilling the adiabaticity condition
$\dot{\om}_{\perp}(t)/\om_{\perp}^2(t)\ll 1$ as long as the decoupling of the transverse degrees of freedom hold.  This allows ultimately to engineer fast modulations of the effective low-dimensional coupling constant $g_{\rm 1D}(t)$ even in a time scale comparable to $\om_{\perp}^{-1}$. Note nonetheless that the excitation of parametric resonances can distort the dynamics \cite{Staliunas04}, as in the experiment \cite{Engels07} were the observation of Faraday waves was reported.
%For Eq. (\ref{effg1d}) and (\ref{1dom2}) to be consistent it is required that
%the dimensionality parameter
%%
%\beqa
%\zeta=N^2a^2/(a_xa_y)\ll 1
%\eeqa
%%
%for all $t>0$ as well. Landau-Zener
%If instead, the dependence of Eq. (6) is considered the non-linearity is found to oscillate periodically in time.

\section{Preserving short-range correlations in time-of-flight measurements}

Time of-flight expansions constitute an essential tool to study quantum correlations in ultracold gases.
Under the assumption of ballistic dynamics it is possible to relate the asymptotic density profile with the momentum distribution of the initial (trapped) state.
The slow power-law decay of the interactions $g_{\rm 1D}(t)=g_{\rm 1D}(0)/[1+\om_{\perp}^2(0)t^2]$, which follows from suddenly switching off the transverse potential,  becomes negligible in the time scale $t\gg \om_{\perp}^{-1}$.
As a result, mean-field interactions are non-negligible in an early stage of the expansion, and blur short-range correlations in a length scale $\delta x\sim c/\om_{\perp}$
%=\xi\mu/\om_{\perp}$,
 where $c$ is the speed of sound. To avoid this, one would like to suppress the role of interactions in a faster time-scale $\tau$.
 %and $\mu$ is the chemical potential \cite{Imambekov09}.
%We further note this functional form of $g_{\rm eff}(t)$ is slowly decaying.
To this aim, consider the reverse engineering of the time-modulation of the transverse trapping frequency that lead to
\beqa
\label{sech}
g_{\rm 1D}(t)=g_{\rm 1D}(0){\rm sech}\frac{t}{\tau},
\eeqa
which satisfies the first two boundary conditions in Eq. (\ref{g0cond}).
Using Eq. (\ref{1dom2}) we find that it can be induced by
%
%\beqa
$
%\label{om4exp}
\om_{\perp}^2(t)=\left(\om_{\perp}(0)^2-\frac{1}{4\tau^2}\right)
{\rm sech}^2\frac{t}{\tau}-\frac{1}{4\tau^2}.
$
For short decay times $\tau$, this trajectory leads to negative values of $\om_{\perp}^2(t)$ associated with purely imaginary frequencies. The physical implementation of $\om_{\perp}^2(t)<0$ requires that the transverse confining potential becomes an expulsive barrier,
pushing the atoms away from the longitudinal axis of the cloud.
This is a general feature of trajectories given by Eq. (\ref{1dom2}) in combination with the first two boundary conditions in Eq. (\ref{g0cond}), and is often required for modulations of the coupling constant $g_{\rm 1D}$ and transverse density in a time scale smaller than  $\om_{\perp}^{-1}$.
% \cite{example}.
For the time dependence in Eq. (\ref{sech}),  whenever $\tau<\frac{1}{\sqrt{2}\om_{\perp}(0)}$, $\om_{\perp}(t)^2<0$ for all $t>0$.  For
$\tau\geq\frac{1}{\sqrt{2}\om_{\perp}(0)}$, $\om_{\perp}(t)^2>0$ at early stages
and becomes negative only after
%
%\beqa
$t_*=\tau {\rm arcsech}\left(\frac{1}{\sqrt{4\om_{\perp}(0)^2\tau^2-1}}\right)$.
%\eeqa
%
Expulsive potentials have already been used in the laboratory, for instance in the study of bright solitons \cite{Khaykovich},
and can be generated in a variety of ways as discussed in \cite{chen10}.
%Depending on the parameters of the trap it might be sufficient to modulate the trapping potential with a positive
%(real) frequency and then switched off at the time of expansion $t_{inv}$.
Moreover, provided that $\ddot{g}_{\rm 1D}(0)\neq 0$, the implementation of the trajectory generally requires sudden jumps of $\om_{\perp}(t)$ as in bang-bang control methods \cite{Salomon}.
% ---------------- FIG. 1 begin ----------------
%
\begin{figure}[t]
\begin{center}
\includegraphics[width=0.8\linewidth]{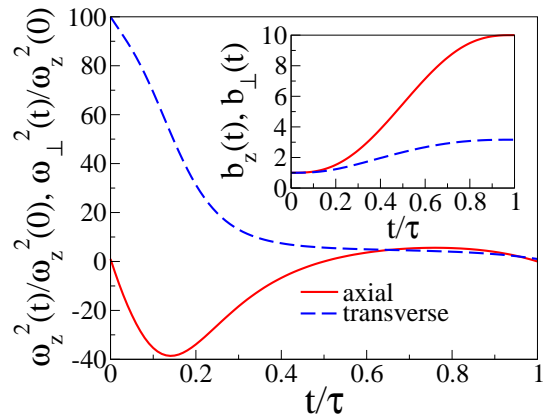}
\end{center}
\caption{\label{fig1} Shortcut to adiabaticity for a quasi-1D  interacting BEC. Outside the Thomas-Fermi regime, the required self-similar dynamics in the axial direction is assisted by a simultaneous modulation of the transverse confinement. The parameters are $\om_z(\tau)=0.01\om_z(0)$ with $b(\tau)=10$, $\om_{\perp}(0)=20\om_z(0)$, $\tau=4\om_z(0)^{-1}$.
The inset shows the smooth evolution of the corresponding scaling factors.
}
\end{figure}

\section{Inducing FFD in quasi-1D Bose-Einstein condensates}
The existence of scalings laws leading to a self-similar dynamics has motivated FFD proposals for superfast expansions providing a shortcut to adiabaticity \cite{chen10,Muga}, as recently demonstrated in the laboratory both with ultracold gases \cite{Labeyrie10} and BECs \cite{Labeyrie10b}.
 When such techniques are extended to quasi-1D BEC, the implementation of a self-similar dynamics generally requires to modulate in time both the axial 
trapping frequency and the coupling constant. As a result, proposals for FFD based on scaling laws become particularly challenging.
In the following, it is shown that a modulation of the transverse trapping frequency can be used to tune the axial effective coupling constant without the need to use a Feschbach resonance.

For the 1D GPE with a time-dependent axial harmonic trap 
$$i\hbar\partial_t\Psi=\big[-\frac{\hbar^2}{2m}\frac{\partial^2}{\partial z^2}+\frac{m\om_z(t)^2z^2}{2}+g_{1D}(t)|\Psi|^2\big]\Psi,$$ 
it is possible to exploit the scaling law for $\Psi=\Psi(z,t)$ in Eq. (\ref{scaling}) for the order parameter, 
with the chemical potential $\mu$ playing the role of the eigen-energy \cite{Muga}. 
The condensate wavefunction evolves self-similarly according to 
\beqa
\label{psiz}
\Psi(z,t)=b_z^{-1/2}e^{i\frac{m\dot{b}_zx^2}{2\hbar b_z}}e^{-i\mu\int_0^t\frac{dt'}{b_z(t')^2}}\Psi\left(\frac{z}{b_z},0\right)
\eeqa 
only if the axial scaling factor $b_z$, obeys the Ermakov equation $\ddot{b}_z+\om_z^2(t)b_z=\om_z^2(0)/b_z^3$ (and $b_z(0)=1$, $\dot{b}_z(0)=0$), and the time-dependent non-linearity of the form $g_{1D}(t)=g_{1D}(0)/b_z(t)$ is implemented. The required time-dependence of $g_{1D}(t)$ could be achieved exploting a Feschbach resonance to tune $g_{3D}$. 
Nonetheless, for a narrow resonance a fast change of the coupling constant -requiring a high control of the external magnetic field- 
can be experimentally challenging \cite{BDZ08}. Alternatively, under isotropic transverse confinement, an axial self-similar dynamics 
under $\om_z(t)$, can be assisted by keeping $g_{3D}$ constant and changing the transverse trapping frequency along the trajectory
\beqa
\label{ombz}
\om_{\perp}^2(t)=\left(\frac{\om_{\perp}(0)}{b_z}\right)^2
+\left(\frac{\dot{b}_z}{2b_z}\right)^2-\frac{\ddot{b}_z}{2b_z},
\eeqa
which implements the required time-dependent non-linear coupling. This trajectory induces a frictionless dynamics in the transverse 
direction (essentially a free harmonic oscilator), which modulates the three dimensional density, and ultimately the effective axial 
non-linearity in the required way $g_{1D}(t)=g_{1D}(0)/b_z(t)$ for the axial dynamics to be self-similar. Once the axial dynamics is 
self-similar, one can engineer $b_z(t)$ to drive an axial FFD.
As a result, to perform an axial FFD of a quasi-1D BEC in a time $\tau$, one can proceed in the following way: i) choose the desired 
initial and final states ($b_z(0)=1$, $b_z(\tau)$) and determine $b(t)$ as in \cite{chen10,Muga}, ii) find the required axial trapping frequency 
\beqa
\label{omz}
\om_z^2(t)=\frac{\om_z(0)^2}{b_z^4}-\frac{\ddot{b}_z}{b_z},
\eeqa 
iii) use Eq. (\ref{ombz})
to derive the required transverse trapping frequency. Implementing both Eqs. (\ref{ombz}) and (\ref{omz}) leads to the self-similar dynamics 
in Eq. (\ref{psiz}) along the designed FFD trajectory $b_z(t)$.
Figure \ref{fig1} shows an instance of these trajectories. Note that thanks to the relation between the axial and transverse scaling factors $b_{\perp}(t)=b_z(t)^{\frac{1}{2}}$, the self-similar dynamics along the $z$-axis can be assisted by a trajectory $\om_{\perp}^2(t)$ involving only real frequencies, without the need to implement a transverse expelling potential but at exceedingly short expansion times $\tau$ or large expansion factors $b(\tau)$.

%Note that either in the axial and transverse direction the Ermakov equation relates the scaling factor with the time-dependnet frequency, and that
%For a desired scaling factor in the axial direction, one can determine the rest of the trajectories following sequence, $b_z(t)\rightarrow \om_z

\section{Nearly sudden quenches}
Another useful tool to probe ultracold gases is the use of a sudden quench of the  interactions. This is the case for the applications mentioned above, as well as for the generation of shock waves and solitons, studies of relaxation dynamics and many other examples. Let us consider a finite time quench between
and initial $g_{1D}(0)=g_i$ and final value $g_{1D}(\tau)=g_f$ of the coupling constant, which approaches the sudden limit as $\tau\rightarrow 0$.  The situation resembles that of FFD.
We require vanishing first and second order derivatives of $g_{1D}(t)$ both at $t=0$ and $ t=\tau$ to avoid transverse excitations  at the end of quench.
It is convenient to consider a polynomial ansatz $g_{1D}(t)=\sum_{l=0}^{5}\alpha_lt^l$, whose coefficients $\{\alpha_l\}$ are completely determined by the boundary conditions, leading to the explicit form of the quench
\beqa
\label{optg1d}
g_{1D}(t)=g_i+[3 s (2 s-5)+10] \gamma  s^3
\eeqa where $\gamma=g_f-g_i$ and $s=t/\tau$.
%
% ---------------- FIG. 1 begin ----------------
%
\begin{figure}[t]
\begin{center}
\includegraphics[width=1.\linewidth]{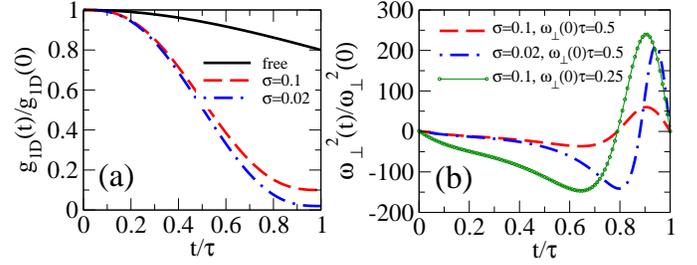}
\end{center}
\caption{\label{fig2} Engineering a quench of the interactions.
a) Decay of the effective non-linear coupling constant $g_{\rm 1D}(t)$ under free expansion in the transverse direction ($\om_{\perp}(t>0)=0$) and different modulations of $\om_{\perp}^2(t)$ engineered according to Eqs. (\ref{1dom2}) and (\ref{optg1d}).
b) Associated trajectories of $\om_{\perp}^2(t)$ for different quenches of $g_{\rm 1D}(t)$.
}
\end{figure}
The computation of the corresponding trajectory $\om_{\perp}(t)$ is straightforward using Eqs. (\ref{1dom2}) and (\ref{optg1d}) but rather lengthy as to be displayed here. Fig. \ref{fig2} shows the solution $\om_{\perp}(t)$ to implement different type of quenches. The required time to achieve a given ratio $\sigma=g_f/g_i<1$  under free evolution is $\tau_{0}=\om_{\perp}^{-1}(0)\sqrt{\gamma^{-1}-1}$, while by implementing $\om_{\perp}(t)$ it is possible to sped up the decay by several orders of magnitude with a moderate modulation of the transverse confinement.
Note that $\sigma=g_f/g_i=b_{\perp}(\tau)^{-2}=\om_{\perp}(0)/\om_{\perp}(\tau)$,  and that the amplitude of the required frequency  scales as $1/[\om_{\perp}(0)\tau]^2$, providing an effective lower bound to the achievable values of $\tau$.

\section{Quasi-2D condensates}
So far we have implicitly focused on a cigar-shape condensate. One can similarly modulate the nonlinearity in a pancake-shaped condensate under strong enough radial confinement such that the dynamics along the most tightly confined direction decouples from that in the BEC plane. Assume an oblate 3D harmonic trap with $\om_x\gg \om_y=\om_z=\om_r$.
% and consider $\epsilon=a_x/\xi\ll 1???$ .
Letting $\Psi(\x,t)=\phi_0(x,t)\psi(y,z,t)$ and $\phi_0(x,t=0)=\exp(-x^2/(2a_x^2)/(\pi^{1/4}\sqrt{a_x})$, using the scaling law for $\phi_0(x,t)$, upon integration of the transverse coordinate, the effective coupling constant for a 2D cloud undergoing a modulation of the transverse confinement is
\beqa
\label{effg2d}
g_{\rm 2D}(t)=g_{\rm 3D}\int dx |\phi_0(x,t)|^4=\frac{g_{\rm 2D}(0)}{b(t)},
\eeqa
where $g_{\rm 2D}(0)=g_{\rm 3D}/(\sqrt{2\pi}a_z)$, with $g_{\rm 2D}(t=0)=g_{\rm 2D}(0)$, $\dot{g}_{\rm 2D}(0)=0$, and as in Eq. (\ref{g0cond}), $\ddot{g}_{\rm 2D}(0)=0$ prevents discontinuous jumps of $\om_{r}(t)$ at $t=0$ from happening.
A given time dependence of $g_{\rm 2D}=g_{\rm 2D}(t)$ follows from a  trajectory
\beqa
\label{2dom2}
\om_x(t)^2=\om_{x}(0)^2\left(\frac{g_{\rm 2D}}{g_{\rm 2D}(0)}\right)^4
+\frac{\ddot{g}_{\rm 2D}}{g_{\rm 2D}}
-2\left(\frac{\dot{g}_{\rm 2D}}{g_{\rm 2D}}\right)^2,
\eeqa
which may imply imaginary frequencies associated with an expulsive potential, as it happens in Eq. (\ref{1dom2}) .
In particular, note that suddenly switching off the transverse potential leads to an essentially linear-in-time decay of the interactions $$g_{\rm 2D}(t)=g_{\rm 2D}(0)/\sqrt{1+\om_x^2(0)t^2}.$$
For a $g_{\rm 2D}(t)$ as in Eq. (\ref{sech}), decaying in a time scale $\tau<\om_x^{-1}$, according to Eq. (\ref{2dom2}) it is found that the required time dependent trajectory  of the control parameter $\om_x^2(t)=\om_x^2(0){\rm sech}^4(t/\tau)-1/\tau^2<0$ requires an expulsive potential for all $t>0$.
We close noticing that in a pancake-shaped cloud, quantum fluctuations are negligible whenever $\Gamma_2=\frac{1}{\sqrt{\pi^3}} \frac{a}{a_x}\ll1$ \cite{HoMa99} and that, should one be interested in tuning the interactions to a larger value $g_{\rm 2D}(\tau)$, the condition becomes $\Gamma_2g_{\rm 2D}(\tau)/g_{\rm 2D}(0)\ll1$.

\section{Discussion and conclusions}
In combination with a spatial dependence of the transverse confining potential, FFD paves the way to control the effective coupling constant both in time and space.
Moreover, further applications can be envisaged such as the preparation of atomic Fock states by many-body atom culling methods \cite{culling}, where starting with a large trapped atomic cloud a controlled increase of the interactions can provide the expelling mechanism for the excess of atoms.

In conclusion, we have presented a scheme to implement  a fast frictionless dynamics of a low-dimensional Bose-Einstein condensate, in which spurious excitations 
are avoided without the need to fulfil adiabaticity constraints. Exploiting the self-similar dynamics in the strongly confined degrees of freedom, we have shown that this can be achieved  by engineering the  modulation of the transverse confinement of the cloud in an elongated trap. As a result, it is possible to tune the amplitude of non-linear interactions in  these systems.
We have further applied the method to preserve short-range correlations in time-of-flight, assist shortcuts to adiabatic expansions in quasi 1D interacting BEC, and implement nearly sudden interaction quenches. More generally, we argue that inverting the equations associated with self-similar scaling laws, allows to determine the trajectory of the control parameter for different processes, and constitute a powerful toolbox for the manipulation of ultracold atoms.

\acknowledgments
It is a pleasure to acknowledge discussions with L. Santos, J. G. Muga, A. Ruschhaupt, X. Chen, and M. D. Girardeau. The author further acknowledges financial support by EPSRC and the European Commission (HIP), as well as the hospitality of the MPIPKS.

\end{document}